\documentclass[journal]{IEEEtran}
\usepackage{multirow}
\usepackage{array}
\usepackage{tabu}
\usepackage{array,booktabs,arydshln,xcolor}
\usepackage{tikz}
\usetikzlibrary{matrix,shapes,arrows,positioning,chains}
\usepackage{amsmath}
 
\newcommand\norm[1]{\left\lVert#1\right\rVert}
\usepackage{amssymb}
\usepackage{enumerate}
\usepackage{algorithm}
\usepackage{amssymb}
\usepackage{textcomp}
\usepackage{commath}
\usepackage{graphicx}
\usepackage{array}
\usepackage{tabu}
\usepackage{array,booktabs,arydshln,xcolor}
\usepackage{tikz}
\usetikzlibrary{matrix,shapes,arrows,positioning,chains}
\usepackage[noend]{algpseudocode}
\usepackage{pifont}
\usepackage{csquotes}
\usepackage{graphicx}
\usepackage{array}
\usepackage{dblfloatfix}
\usepackage{pgfplots}
\usepackage{tikzscale}

\usepackage{csquotes}
\usepackage{float}
\usepackage{cite}
\usepackage{breqn}
\usepackage{blox}
\usetikzlibrary{shapes,arrows}

\ifCLASSINFOpdf  
\else
\fi
\begin{document}
\title{Robust Adaptive Neural Network Control of Time-Varying State Constrained Nonlinear Systems
\thanks{The authors are with the Department of Electrical Engineering,
Indian Institute of Technology Kanpur, Kanpur 208016, India (e-mail:
pmpankajjec@gmail.com; nishchal.iitk@gmail.com.}}

\author{Pankaj K. Mishra
        and Nishchal K Verma}

\maketitle

\begin{abstract}
 This paper deals with the tracking control problem for a  very simple class of unknown nonlinear systems. In this paper, we  presents a design strategy for tracking control of time-varying state constrained nonlinear systems in an adaptive framework. The controller is designed using the backstepping method. While designing it, Barrier Lyapunov Function (BLF) is used so that the state variables do not contravene its constraints. In order to cope with the unknown dynamics of the system, an online approximator is designed using a neural network with a novel adaptive law for its weight update.  To make the controller robust and computationally inexpensive, a disturbance observer is proposed to cope with the disturbance along with neural network approximation error and the time derivative of virtual control input. The effectiveness of the proposed approach is demonstrated through a simulation study.
\end{abstract}


\IEEEpeerreviewmaketitle

\section{Introduction}

Recently, researchers in the field of nonlinear control systems have made significant efforts to address the issue of system state and output stability. However, in everyday life, numerous uncertain dynamic systems have constraints such as performance, saturation, physical stoppages, and safety specifications. Constraints are ineludible for such systems when designing controllers in real-time. In practical systems, constraints can be static or dynamic, and their upper and lower bounds can be symmetric or asymmetric. The Barrier Lyapunov Function (BLF) has been widely used in the literature to deal with such systems. In order to design a controller for a system with static symmetric or asymmetric constraints, \cite{TEE2009918} provides a nice integration of BLF with the well-known backstepping technique. Other than the BLF-based technique, other efforts have been undertaken by academia and industry to design a controller for the constrained system, including error transformation and model predictive control (MPC).
In error transformation, the application of tangent hyperbolic in a prescribed function may lead to a singularity problem, and exorbitant control input may violate prescribed control performance, leading to instability. In MPC, linear and nonlinear system constraints are addressed by solving a finite horizon open-loop optimal control problem \cite{MAYNE2000789}. Most optimal control and MPC rely on numerical, computationally intensive algorithms to solve control problems \cite{kirk_2016}. BLF has been studied for the controller design of constrained systems because it easily handles unknown system dynamics, uncertainties, and disturbances by integrating robust adaptive backstepping or sliding mode control. In \cite{8668697,8952870,LI2018444, 8353903, 8290541,8879661}, authors have used BLF to design controller for static state constrained nonlinear systems. Further, in \cite{8732686, XI2019108,SONG2018314,8948553, 8603838,8815865,8759967}, authors have designed controller for time-varying state constrained nonlinear systems, however design are not robust.\par Motivated by the aforementioned works, the contributions of this paper are listed below.

 \begin	{enumerate}
\item A novel adaptive law for neural networks (NN) is designed to deal with unknown dynamics of the systems.
\item  To deal with the uncertainties such as disturbance, approximation error and explosion of derivative of virtual control law in the backstepping design, a novel disturbance observer has been proposed.
\item Further, to deal with unknown control gain a novel controller has been proposed using Nussbaum gain.
 \end{enumerate}

The paper is organized as follows. In Section \ref{s2.2}, we present the system description and problem statement.  This section also presents some assumptions, definition, and lemmas for the stability analysis of the system.  Section \ref{s2.3} discusses the construction of NN for the approximation of unknown terms involved in the design; Section \ref{s2.4} consists of two subsections. Subsection \ref{s2.41} discusses the design of a disturbance observer for the robustness of the system, and Subsection \ref{s2.42} discusses the steps to design an adaptive controller using the backstepping technique. Section \ref{s2.5} discusses the theorem for the boundedness of the signals in the closed-loop system. Section \ref{s2.6} illustrates the proposed methodology using the simulation examples. Finally, Section \ref{s2.7} concludes the paper.

\section{System Description and Problem Statement}\label{s2.2}
Consider a class of  SISO nonlinear systems shown below
\begin{equation}\label{sys1}
\begin{split}
\dot x_i&=x_{i+1}+d_i\left(t\right)~~~~\forall i \in \{1, \hdots, n-1\}\\
\dot x_n&=f\left(x\right)+\beta u+ d_n\left(t\right)\\
y&=x_1
\end{split}
\end{equation}
where $x_i\in \mathbb{R}$,  $\forall i \in \mathbb{N}_n$, $u\in \mathbb{R}$ and $y\in \mathbb{R}$ are the $i^{th}$ state,  the control input, and the output of the  system, respectively; $f\left(x\right)$ $\in \mathbb{R}$ is smooth  unknown nonlinear functions  and $\beta$ is unknown control coefficient;  $d_i \in \mathbb{R} , \forall i \in \mathbb{N}_n$ are unknown time-varying bounded disturbance. In this study, states are considered to be constrained such that, $\abs{x_i}<\Psi_i(t)$, where $\Psi_i \in \mathbb{R}$ is a known time varying state constraint on the state variable.\par
{\textit{Problem Statement:}} The goal of this paper is to design a NN-based adaptive controller for  (\ref{sys1}) such that (i) output $y$ tracks the desired output $y_d$; (ii) all the closed-loop signals are guaranteed to be bounded;  and (iii) all the system states do not contravene their state constraints.\par
Following are the assumptions, definition and lemmas, which will be needed to achieve the control objective.\par
\textit{Assumption 1 \cite{LIU2017143}:}  The control coefficient $\beta\neq 0$.\par
\textit{Assumption 2 \cite{1220768}:} The unknown time-varying $\text{disturbance}$ $d_{i}(t)$ is bounded and there exist some positive constant $d_0$ such that $\abs {\dot d_{i}(t)} \le d_0$ $\forall i \in \mathbb{N}_n$.\par
\textit{Assumption 3 \cite{7429795, LIU2017143, 7571100}.} The first $n${th} time derivative of desired output $y_d$ is bounded $\forall t \in [0,\infty)$.\par
\textit{Remark 1:} For the computation of time derivative of virtual control input, we need time derivative of desired output in each step of backstepping based scheme, so its availability and boundedness is a must. However, here we have relaxed the availability by estimating the time derivative of virtual control input using the disturbance observer. \par
\textit{Assumption 4 \cite{7786843}.} The time-varying  symmetric state constraint  $\Psi_i(t) \in \mathbb{R}$ is bounded $\forall t \in [0,\infty)$.\par
\textit{Definition 1  \cite{Nussbaum1983}} : The  function $\mathcal{N}(\zeta)$ is said to be Nussbaum, if it holds the following property:\par
	\begin{equation}
	\begin{split}
	\underset{s \rightarrow \infty}{\lim} \sup \frac{1}{s}\int_{0}^s \mathcal{N}(\zeta)d\zeta = +\infty\\
	\underset{s \rightarrow \infty}{\lim} \inf \frac{1}{s}\int_{0}^{s} \mathcal{N}(\zeta)d\zeta = -\infty.
	\end{split}
	\end{equation}
	There are several functions that can be considered Nussbaum functions, including  $e^{\zeta^2}\cos((\pi/2) \zeta)$ and $\zeta^2cos(\zeta)$. In this  paper, we have used $\mathcal{N}(\zeta)=\zeta^2cos(\zeta)$  as a Nussbaum function. \par
\textit{Lemma 1 \cite{1220768}:} Let $\mathcal{V}(t)\ge 0$ and $\zeta(t)$ be  smooth functions defined on $[0, t_f)$ and  $\mathcal{N}(\zeta(t))$  be an even smooth Nussbaum function. If the following inequality holds: \par
	\begin{align}
	\mathcal{V}(t)\le \kappa_1+e^{-\kappa_2t}\int_{0}^{t}\left({\beta_0\mathcal{N}(\zeta)}+1\right)\dot\zeta e^{\kappa_2 \tau}d\tau
	\end{align}
	where $\kappa_1$ and $\kappa_2$ are positive constant, and $\beta_0$ is a non-zero constant, then $\mathcal{V}(t)$, $\zeta(t)$ and $\int_{0}^{t}{\beta_0\mathcal{N}(\zeta)}\dot\zeta d\tau$ are bounded on $[0, t_f)$.\par
\textit{Lemma 2 \cite{5499019}:} For any $z$ in the interval $\abs{z} < \abs{\psi}$, where  $\psi \in \mathbb{R}$, we have 
\begin{equation}
 \log\frac{\psi^2}{\psi^2-z^2}<\frac{z^2}{\psi^2-z^2}.
\end{equation}
\section{NN Approximation}\label{s2.3}
The function $f(x)$ is not known in the system (\ref{sys1}). This section will look at an online approximation strategy for dealing with an unknown function. For this, Radial Basis Function (RBF) NN is used. It is well known that using the universal approximation property of RBF  NN, we can approximate any unknown continuous function. The RBF NN used here has $l$ number of hidden neurons and a output. The output of the network $O_{NN}(\theta,\bar z) \in \mathbb{R}$ is given by
\begin{equation}
O_{NN}(\theta,\bar z)=\theta^T\varphi(\bar z) \quad 
\end{equation}
where the vector $\bar z= [x_1, \hdots, x_n]^T$ is the input of the  NN, $ \theta=[\theta_i, \ldots, \theta_l] \in \mathbb{R}^{l}$ is the weight vector, $\varphi(\bar z) \in \mathbb{R}^l$ is a basis vector of RBF NN with a set of suitably chosen Gaussian basis function $(\varphi_i \in \mathbb{R}$, $\forall i\in \mathbb{N}_l)$ defined on a compact set $\Omega _ {\bar z}$, such that  $\varphi=[\varphi_1(\bar z),\ldots, \varphi_l(\bar z)]^T$ and 
\begin{equation}
\varphi_i(\bar z)=exp\left(\frac{-\norm{\bar z-c_i}^2}{b_i}\right) \qquad \forall i \in \mathbb{N}_l
\end{equation}
where $c_i \in \Omega_{\bar z}$ is the centre of receptive field and $b_i \in \mathbb{R}$ is the width of Gaussian function. From the definition of $\varphi_i(\bar z)$, we find that it is bounded. Let say $\bar \varphi$ be the upper bound of $\varphi_i(\bar z)$ then
	\begin{align}\label{2bnd1}
	\norm{\varphi(\bar z)}\le\bar \varphi.
	\end{align}
Assuming that an ideal weight vector $\theta^{*}=[\theta^*_{1}, \ldots, \theta^*_{l}] \in \mathbb{R}^{l}$ exists,  such that
\begin{equation}
f(x)=\theta^{*T}\varphi(\bar z)+\epsilon(\bar z) \quad  \label{2fx}
\end{equation}
where, $\theta^*$ and $\epsilon$ are   ideal weight vector and  approximation error respectively.\par
\textit{Assumption 5.} The approximation error vector $\epsilon$ is bounded and  $\abs{\epsilon}\le \bar \epsilon $ for some positive constant $\bar \epsilon $.\par
The ideal weight vector $\theta  ^*$ is defined as follows
\begin{equation}
\begin{split}
\theta^*=\arg \underset { \theta^*\in \mathbb{R}^{l }}{\min}\left\lbrace  {\sup} \left( O_{NN}(\theta^*,\bar z)-  O_{NN}(\theta,\bar z)\right)\right\rbrace.
\end{split}
\end{equation}
Using (\ref{2fx}), the system (\ref{sys1}) can be rewritten as
\begin{align}
\dot x_i&=x_{i+1}+d_i\left(t\right)~~~~~\forall i \in \mathbb{N}_{n-1}\label{2si}\\
\dot x_n&=\theta^{*T}\varphi(\bar z)+\epsilon(\bar z)+\beta u+ d_n\left(t\right).\label{2fi}
\end{align}
The ideal weight matrix $\theta^{*}$ above is not known and therefore needs to be estimated. Let $ \hat \theta =[\hat \theta_{1}, \ldots, \hat \theta_{l}]  \in \mathbb{R}^{l}$ be the estimate of ideal weight  matrix  $ \theta^*$ such that
\begin{equation}
\hat f(x)=\hat \theta^T \varphi(\bar z)
\end{equation}
where $\hat f(x)$ is an approximation of the unknown nonlinear function $f(x)$. The next steps of controller design have been presented in the following section.
\section{Robust Adaptive Backstepping Controller Design}\label{s2.4}
Let  $z=[z_{1}, \ldots , z_{n}]^T$, $v=[v_{1}, \ldots, v_{n-1}]^T$, and $v_0=x_{1d} $ be an error vector, virtual control input vector, and desired output vector respectively. The error vector elements are defined as follows
\begin{equation}\label{2ervr}
\begin{split}
z_{i}&=x_{i}-v_{i-1} \qquad \forall i \in \mathbb{N}_{n}.\\
\end{split}
\end{equation}
\textit{Note:} To maintain the uniformity in the expression for the error variables, it is common practice to denote the desired output with a symbol,  similar to virtual control input with  $0$ in subscript.\par
On differentiating (\ref{2ervr}) with respect to time and using (\ref{2si}), $\forall i \in \mathbb{N}_{n-1}$  the error dynamics is
\begin{align}
\dot z_{i}&=x_{i+1}+d_{i}(t)-\dot v_{i-1},  \label{2ed}
\end{align}
and taking the time derivative of (\ref{2ervr}) and using (\ref{2fi}) for $i=n $, we have error dynamics
\begin{align}
\dot z_{n}&=\theta^{*T}\varphi(\bar z)+\epsilon(\bar z)+\beta u+ d_n\left(t\right) -\dot v_{n-1}. \label{2edn}
\end{align}

\subsection{Disturbance Observer}\label{s2.41}
The calculation of the derivative of virtual control input is a major computing step in backstepping-based controller design. The derivative of this control input must be estimated. The disturbance observer is designed to have an estimate including the unknown disturbance.\par
The observer variable $\varepsilon=[\varepsilon_{1}, \ldots, \varepsilon_{n}]^T \in \mathbb{R}^n$ is defined as
\begin{align}
\varepsilon _{i}&=d_{i}(t)-\dot v_{i-1}, \text{\space} \forall i\in \mathbb{N}_{n-1}\label{2ov1}\\
\varepsilon_{n}&=\epsilon(\bar z)+ (\beta-1)u + d_n\left(t\right) -\dot v_{n-1}.  \label{2ovn}
\end{align}
\textit{Assumption 6.} The observer variable $\varepsilon_i$, to be estimated, is bounded and $\forall i \in \mathbb{N}_n$ there exists a positive constant $\rho_i$ such that its derivative $\abs{\dot \varepsilon_i}\le\bar \varepsilon_i$.\par
The error dynamics (\ref{2ed}) and (\ref{2edn}) can be expressed  using (\ref{2ov1}) and (\ref{2ovn}) respectively as
\begin{align}
\dot z_{i}&=x_{i+1}+\varepsilon _{i}, \text{\space} \forall i \in \mathbb{N}_{n-1}  \label{2ed1}\\
\dot z_{n}&=\theta^{*T}\varphi(\bar z)+  u +\varepsilon _{n}. \label{2ed2}
\end{align}
To estimate the observer variable in (\ref{2ov1}) and (\ref{2ovn}) an auxiliary system  is introduced. It is defined as 
\begin{align}
\eth _i&=\varepsilon _i-k_{\varepsilon _i}z_i, \quad \forall i\in \mathbb{N}_n\label{2eth}
\end{align}
{where $k_{\varepsilon _i}$ is an observer gain.\par
Using (\ref{2ed1}) and (\ref{2ed2}), we can rewrite the  dynamics of auxiliary  system (\ref{2eth}) as
\begin{align}
\dot \eth _i&=\dot\varepsilon _i-k_{\varepsilon _i}(x_{i+1}+\varepsilon _{i}), \quad \forall i\in \mathbb{N}_{n-1} \label{2deth1}\\
\dot \eth _n&=\dot\varepsilon _n-k_{\varepsilon _n}(\theta^{*T}\varphi(\bar z)+ u +\varepsilon _{n}). \label{2deth2}
\end{align}
To estimate the auxiliary system, the observer dynamics is proposed as 
\begin{align}
\dot {\hat \eth}_i&=-k_{\varepsilon _i}(x_{i+1}+\hat\varepsilon _{i})~~\text{and} \label{2ethj}\\
\dot {\hat \eth}_n&=-k_{\varepsilon _n}(\hat \theta^{T}\varphi(\bar z)+ u +\hat\varepsilon _{n}).\label{2ethn}
\end{align}
Using (\ref{2eth}), the estimate of observer variables (\ref{2ov1}) and (\ref{2ovn}) can be obtained as 
\begin{align} \label{2varhat}
\hat \varepsilon_i=\hat \eth_i+k_{\varepsilon _i}z_i, \quad \forall i\in \mathbb{N}_{n}. 
\end{align}

Using (\ref{2eth}) and (\ref{2varhat}), the estimation error $\tilde{\eth}_i$ of  the auxiliary system  can be written as
\begin{align} \label{2auxd}
\tilde{\eth}_i=\eth_i-\hat \eth_i=\tilde \varepsilon_i, \quad \forall i\in \mathbb{N}_{n}.
\end{align}
Subtracting (\ref{2ethj}) and (\ref{2ethn}) from (\ref{2deth1}) and (\ref{2deth2}) respectively, and using (\ref{2auxd}),  the observer error dynamics for the auxiliary system becomes
\begin{align}
\dot {\tilde \eth}_i&=\dot {\tilde \varepsilon}_i=\dot{\varepsilon} _i-k_{\varepsilon _i}\tilde\varepsilon _{i}, \quad \forall i\in \mathbb{N}_{n-1}\label{2oal}\\
\dot {\tilde \eth}_n&=\dot {\tilde \varepsilon}_n=\dot{\varepsilon} _n-k_{\varepsilon _n}(-\tilde\theta^{T}\varphi(\bar z)+\tilde\varepsilon _{n}),\label{2oaln}
\end{align}
where $\tilde \theta=\hat \theta -\theta^*$ and $\tilde \varepsilon_i=\varepsilon_i- \hat \varepsilon,$ $\forall i \in \mathbb{N}_n$.

\subsection{Controller Design and Stability Analysis}\label{s2.42}
To begin, first we  will define $n$ BLF for the  $n$ states of the system (\ref{sys1}) as well as their time derivative will be calculated.\par
Let $\mathcal{L}_i$, $i=1, \hdots, n$ be a BLF and defined as
\begin{align}\label{2L}
\mathcal{L}_i=\frac{1}{2}\log\frac{\psi_i^2( t)}{\psi_i^2( t)-z_i^2},
\end{align}
where $\psi_i( t)$ is a constraint on error variable $z_i$, which will be defined later.
Taking the time derivative of (\ref{2L}), we have
\begin{align}\label{2ldot}
\dot {\mathcal{L}}_i={\mathcal{Q}}_i\left(\dot z_i-\frac{z_i}{\psi_i}\dot\psi_i\right),
\end{align}
\vspace{-0.5cm}
\begin{align}\label{2Qi}
\text{where~} {\mathcal{Q}}_i=\frac{z_i}{\psi_i^2-z_i^2}. 
\end{align}
Following (\ref{2ed}) and  substituting $z_i=x_i-v_{i-1}$ in  (\ref{2ldot}), we have
\begin{align}\label{2ldot1}
\dot {\mathcal{L}}_i={\mathcal{Q}}_i\left(\dot x_i-\dot v_{i-1}-\frac{z_i}{\psi_i}\dot\psi_i\right).
\end{align}
The design steps of controller are as follows:\par
\textit{Step 1:} Consider a Lyapunov function $\mathcal{V}_1$, as
\begin{align}\label{2v1}
{\mathcal{V}}_1=\mathcal{L}_1+\frac{1}{2}\tilde{\varepsilon}^2_1.
\end{align}
Taking the time derivative of (\ref{2v1})  and using (\ref{2ldot}),  we have
\begin{align}\label{2v1d}
\dot{\mathcal{V}}_1={\mathcal{Q}}_1\left(\dot z_1-\frac{z_1}{\psi_1}\dot\psi_1\right)+{\tilde{\varepsilon}}_1\dot {\tilde{\varepsilon}}_1.
\end{align}
On substituting (\ref{2ed1}) for $i=1$ in (\ref{2v1d}), we have
\begin{align}\label{2v1dd}
\dot{\mathcal{V}}_1={\mathcal{Q}}_1\left( x_{2}+\varepsilon _{1}-\frac{z_1}{\psi_1}\dot\psi_1\right)+{\tilde{\varepsilon}}_1\dot {\tilde{\varepsilon}}_1.
\end{align}

On substituting (\ref{2oal}) for $i=1$ in (\ref{2v1dd}), we have
\begin{align}\label{22v1ddd}
\dot{\mathcal{V}}_1={\mathcal{Q}}_1 x_{2}+{\mathcal{Q}}_1\varepsilon _{1}-{\mathcal{Q}}_1\frac{z_1}{\psi_1}\dot\psi_1+{\tilde{\varepsilon}}_1\dot{\varepsilon} _1-k_{\varepsilon _1}\tilde\varepsilon _{1}^2.
\end{align}

Following (\ref{2ervr}), and substituting  $x_{2}=z_{2}+v_{1}$ in (\ref{22v1ddd}) leads to
\begin{align}\label{2v1ddd}
\dot{\mathcal{V}}_1={\mathcal{Q}}_1 z_{2}+{\mathcal{Q}}_1 v_{1}+{\mathcal{Q}}_1\varepsilon _{1}-{\mathcal{Q}}_1\frac{z_1}{\psi_1}\dot\psi_1+{\tilde{\varepsilon}}_1\dot{\varepsilon} _1-k_{\varepsilon _1}\tilde\varepsilon _{1}^2.
\end{align}

Choose the virtual controller $v_1$  as
\begin{align}
v_1&=\mathcal{N}_1(\zeta_1)\alpha_1,~\text{where} \label{2vv1}\\
\dot \zeta_1&=\mathcal{Q}_1\alpha_1,~\text{and}\label{2g1}\\
\alpha_1&=k_1z_1+\hat\varepsilon_1+{\mathcal{Q}_1}-\frac{z_1}{\psi_1}{\dot \psi_1},\label{2a1}
\end{align}
and the design parameter $k_1>0$.\par
Using  (\ref{2vv1})-(\ref{2a1}), (\ref{2v1ddd}) becomes
\begin{align}\label{2v1dddd}
\dot {\mathcal{V}}_1&=- k_1\mathcal{Q}_1z_1+\mathcal{Q}_1z_{2}+ \mathcal{N}_1(\zeta_1)\dot \zeta_1+\dot \zeta_1\notag\\
&\quad +\mathcal{Q}_1{\tilde{\varepsilon}_1}+{\tilde{\varepsilon}}_1\dot{\varepsilon} _1-k_{\varepsilon _1}\tilde\varepsilon _{1}^2-{\mathcal{Q}_1^2}.
\end{align}
For further analysis, we need few inequality relations. They are as follows
\begin{enumerate}[i)]
	\item First term of (\ref{2v1dddd}), i.e.,  $k_1\mathcal{Q}_1z_1$:\\
	Following (\ref{2Qi}) and using  Lemma 2,  we have
	\begin{align}
	- \frac{1}{2}\mathcal{Q}_1z_1=-\frac{1}{2}\frac{z_1^2}{\psi_1^2-z_1^2}\le- \frac{1}{2}\log \frac{\psi_1^2}{\psi_1^2-z_1^2} \label{2inl}.
	\end{align}
	Multiplying  (\ref{2inl}) on both sides by $2k_1$, we have  
	\begin{align}
	- k_1\mathcal{Q}_1z_1\le-2k_1 \mathcal{L}_1 \label{2Fi}.
	\end{align}
\end{enumerate}
\begin{enumerate}[i)]
	\setcounter{enumi}{1}
	\item   Second term of (\ref{2v1dddd}), i.e., $\mathcal{Q}_1z_{2}$ :\\
	Using the Young's inequality, we have
	\begin{align}
	    \mathcal{Q}_1z_{2}\le\frac{\mathcal{Q}_1^2}{2}+\frac{z_2^2}{2}
	\end{align}
\end{enumerate}
\begin{enumerate}[i)]
	\setcounter{enumi}{2}
	\item Fifth, sixth, and seventh term of (\ref{2v1dddd}), i.e., $\mathcal{Q}_1{\tilde{\varepsilon}_1}+{\tilde{\varepsilon}}_1\dot{\varepsilon} _1-k_{\varepsilon _1}\tilde\varepsilon _{1}^2$.
\end{enumerate}
Following  Assumption 6,  and  using Young's inequality, we have
\begin{align}\label{2Ti}
\mathcal{Q}_1{\tilde{\varepsilon}_1}+{\tilde{\varepsilon}}_1\dot{\varepsilon} _1-k_{\varepsilon _1}\tilde\varepsilon _{1}^2 &\le\frac{\tilde \varepsilon_1^2}{2}+\frac{\mathcal{Q}_1^2}{2}+\frac{\tilde \varepsilon_1^2}{2}+\frac{{\bar \varepsilon}_1^2}{2}-k_{\varepsilon_1}{\tilde{\varepsilon}}^2_1,\notag\\
&=-\tilde \varepsilon_1^2\left(k_{\varepsilon_1}-1\right)+\frac{\mathcal{Q}_1^2}{2}+\frac{{\bar \varepsilon}_1^2}{2}.
\end{align}
Using  the  inequalities (\ref{2Fi}) and (\ref{2Ti}),  in  (\ref{2v1dddd}), we have 
\begin{IEEEeqnarray}{rcl}
\dot {\mathcal{V}}_1&\le&   \mathcal{N}_1(\zeta_1)\dot \zeta_1+\dot \zeta_1 -2k_1{\mathcal{L}}_1 +z_{2}^2\nonumber\\
&\quad&-\tilde \varepsilon_1^2\left(k_{\varepsilon_1}-1\right)+\varrho_1. \IEEEeqnarraynumspace \label{2ldot11}
\end{IEEEeqnarray}
where $\varrho_1=\frac{{\bar \varepsilon}_1^2}{2}$.\par
Equation (\ref{2ldot11}) can be further written as
	\begin{align}\label{2ldot12}
	\dot{\mathcal{V}}_1&\le-\mu_1\mathcal{V}_1+  \mathcal{N}_1(\zeta_1)\dot \zeta_1+\dot \zeta_1 +z_{2}^2+\varrho_1,
	\end{align}
	where  $\mu_1=\min\left(2k_1,2\left(k_{\varepsilon_1}-1\right)\right)$.\par
In the  decoupled backstepping design, we will  seek for the boundedness of $z_2$ in  the next step of the  design rather than cancellation of  $z_2^2$.\par
	Multiplying both sides of (\ref{2ldot12}) by $e^{\mu_1 t}$, we have
	\begin{align}\label{2ldot13}
	\frac{d(\mathcal{V}_1(t)e^{\mu_1 t})}{dt}\le\left({\mathcal{N}}_1(\zeta_1)\dot\zeta_1+\dot\zeta_1+z_{2}^2+\varrho_1\right)e^{\mu_1 t}. 
	\end{align}
	Integrating (\ref{2ldot13}) over $\left[0,t\right]$, gives
	\begin{align}
	e^{\mu_1 t}\mathcal{V}_1(t)&\le \mathcal{V}_1(0) + \int_{0}^{t}{ \left({\mathcal{N}}_1(\zeta_1)+1\right) \dot\zeta_1e^{\mu_1\tau}d\tau}\notag\\
	&\quad +\int_{0}^{t}{z_{2}^2e^{\mu_1 \tau}d\tau}+\frac{\varrho_1 e^{\mu_1 t}}{\mu_1}-\frac{\varrho_1}{\mu_1} \label{2ldot14}.
	\end{align}
	On multiplying both sides of (\ref{2ldot14}) by $e^{-\mu_1 t}$, we have
	\begin{align}
	\mathcal{V}_1(t)&\le e^{-\mu_1 t}\mathcal{V}_1(0) + e^{-\mu_1 t}\int_{0}^{t}{ \left({\mathcal{N}}_1(\zeta_1)+1\right) \dot\zeta_1e^{\mu_1\tau}d\tau}\notag\\
	&\quad +e^{-\mu_1 t}\int_{0}^{t}{z_{2}^2e^{\mu_1 \tau}d\tau}+\frac{\varrho_1 }{\mu_1}-\frac{\varrho_1e^{-\mu_1 t}}{\mu_1}. \label{2ldot15}
	\end{align}
	Since, $0<e^{-\mu_1 t}\le1$, we can write (\ref{2ldot15}) as
	\begin{align}
	\mathcal{V}_1(t)&\le \mathcal{V}_1(0) + e^{-\mu_1 t}\int_{0}^{t}{ \left({\mathcal{N}}_1(\zeta_1)+1\right) \dot\zeta_1e^{\mu_1\tau}d\tau}\notag\\
	&\quad +e^{-\mu_1 t}\int_{0}^{t}{z_{2}^2e^{\mu_1 \tau}d\tau}+\frac{\varrho_1 }{\mu_1}-\frac{\varrho_1e^{-\mu_1 t}}{\mu_1}. \label{2ldot166}
	\end{align}
	We can rewrite (\ref{2ldot166}) as
	\begin{align}
	\mathcal{V}_1(t)&\le \mathcal{V}_1(0) + e^{-\mu_1 t}\int_{0}^{t}{ \left({\mathcal{N}}_1(\zeta_1)+1\right) \dot\zeta_1e^{\mu_1\tau}d\tau}\notag\\
	&\quad +e^{-\mu_1 t}\int_{0}^{t}{z_{2}^2e^{\mu_1 \tau}d\tau}+\frac{\varrho_1 }{\mu_1}. \label{2ldot1666}
	\end{align}
In (\ref{2ldot166}), if there would have been no extra term, i.e.  $e^{-\mu_1 t}\int_{0}^{t}{z_{2}^2e^{\mu_1 \tau}d\tau}$, then using Lemma 1, we may have shown that $\mathcal{V}_1(t),\zeta_1$ and $z_1, \hat \varepsilon_1$ are all uniformly ultimately bounded. However, if we can show  $z_2$ is bounded, then using the following relation 
	\begin{align}
	e^{-\mu_1 t}\int_{0}^{t}{z_{2}^2e^{\mu_1 \tau}d\tau}&\le e^{-\mu_1 t}\underset{\tau\in[0,t]}{\sup} z_{2}^2\int_{0}^{t}{e^{\mu_1 \tau}d\tau} \notag\\
	&\le \frac{\underset{\tau\in[0,t]}{\sup} z_{2}^2}{\mu_1}, \label{2bound1}
	\end{align}
	we can say that $e^{-\mu_1 t}\int_{0}^{t}{z_{2}^2e^{\mu_1 \tau}d\tau}$ is bounded. Consequently using Lemma 1, we will be able to show $\mathcal{V}_1(t),\zeta_1$ and $z_1,\hat \varepsilon_1$ are also bounded. Again to show $z_2$ is bounded, we need to follow similar steps. The process will be recursive until we do not  have  $z_{i+1}^2$ in the derivative of Lyapunov function.	\par
\textit{Step i} \textit{$(i=2, \hdots,n-1)$}: Consider a Lyapunov function 
\begin{align}\label{2vi}
{\mathcal{V}}_i=\mathcal{L}_i+\frac{1}{2}\tilde{\varepsilon}^2_i.
\end{align}

Taking the time derivative of (\ref{2vi})  and using (\ref{2ldot}),  (\ref{2vi}) becomes
\begin{align}\label{2vid}
\dot{\mathcal{V}}_i={\mathcal{Q}}_i\left(\dot z_i-\frac{z_i}{\psi_i}\dot\psi_i\right)+{\tilde{\varepsilon}}_i\dot {\tilde{\varepsilon}}_i.
\end{align}
On substituting (\ref{2ed1})  in (\ref{2vid}), we have
\begin{align}\label{22vidd}
\dot{\mathcal{V}}_i={\mathcal{Q}}_i\left( x_{i+1}+\varepsilon _{i}-\frac{z_i}{\psi_i}\dot\psi_i\right)+{\tilde{\varepsilon}}_i\dot {\tilde{\varepsilon}}_i.
\end{align}

On substituting (\ref{2oal}) in (\ref{22vidd}), we have
\begin{align}\label{2vidd}
\dot{\mathcal{V}}_i={\mathcal{Q}}_i x_{i+1}+{\mathcal{Q}}_i\varepsilon _{i}-{\mathcal{Q}}_i\frac{z_i}{\psi_i}\dot\psi_i+{\tilde{\varepsilon}}_i\dot{\varepsilon} _i-k_{\varepsilon _i}\tilde\varepsilon _{i}^2.
\end{align}

Following (\ref{2ervr}), and substituting  $x_{i}=z_{i}+v_{i-1}$ in (\ref{2vidd}) leads to
\begin{align}\label{2viddd}
\dot{\mathcal{V}}_i={\mathcal{Q}}_i z_{i+1}+{\mathcal{Q}}_i v_{i}+{\mathcal{Q}}_i\varepsilon _{i}-{\mathcal{Q}}_i\frac{z_i}{\psi_i}\dot\psi_i+{\tilde{\varepsilon}}_i\dot{\varepsilon} _i-k_{\varepsilon _i}\tilde\varepsilon _{i}^2.
\end{align}

Choose the virtual controller $v_i$  as
\begin{align}
v_i&=\mathcal{N}_i(\zeta_i)\alpha_i,~\text{where} \label{2vvi}\\
\dot \zeta_i&=\mathcal{Q}_i\alpha_i,~\text{and}\label{2gi}\\
\alpha_i&=k_iz_i+\hat\varepsilon_i+\mathcal{Q}_i-\frac{z_i}{\psi_i}{\dot \psi_i},\label{2ai}
\end{align}
where $k_i>0$ is a design parameter.\par
Using  (\ref{2vvi})-(\ref{2ai}), in (\ref{2viddd}) and 	following the same procedure as step 1, we have
	\begin{align}
	\mathcal{V}_i(t)&\le \mathcal{V}_i(0) + e^{-\mu_i t}\int_{0}^{t}{ \left({\mathcal{N}}_i(\zeta_i)+1\right) \dot\zeta_ie^{\mu_i\tau}d\tau}\notag\\
	&\quad +e^{-\mu_i t}\int_{0}^{t}{z_{i+1}^2e^{\mu_i \tau}d\tau}+\frac{\varrho_i }{\mu_i} \label{2ldot16}
	\end{align}
	where $\mu_i=\min\left(2k_i,2\left(k_{\varepsilon_i}-1\right)\right)$, $\varrho_i=\frac{{\bar \varepsilon}_i^2}{2}$ and 
	\begin{align}
	e^{-\mu_i t}\int_{0}^{t}{z_{i+1}^2e^{\mu_i \tau}d\tau}&\le e^{-\mu_i t}\underset{\tau\in[0,t]}{\sup} z_{i+1}^2\int_{0}^{t}{e^{\mu_i \tau}d\tau} \notag\\
	&\le \frac{\underset{\tau\in[0,t]}{\sup} z_{i+1}^2}{\mu_i}. \label{2boundi}
	\end{align}
 Similar to previous discussion in step 1, we can apply Lemma 1 to show $\mathcal{V}_i(t),\zeta_i$ and $z_i,\hat \varepsilon_i$ are all uniformly ultimately bounded, provided $z_{i+1}$ is bounded.

\noindent\textit{Step n} }: Consider a Lyapunov function 
\begin{align}\label{2vn}
{\mathcal{V}}_n=\mathcal{L}_n+\frac{1}{2}\tilde{\varepsilon}^2_n+\frac{1}{2\lambda}\Tilde{\theta}^T\Tilde{\theta}.
\end{align}

Taking the time derivative of (\ref{2vn})  and using (\ref{2ldot}),  (\ref{2vn}) becomes
\begin{align}\label{2vnd}
\dot{\mathcal{V}}_n={\mathcal{Q}}_n\left(\dot z_n-\frac{z_n}{\psi_n}\dot\psi_n\right)+{\tilde{\varepsilon}}_n\dot {\tilde{\varepsilon}}_n+\frac{1}{\lambda}\Tilde{\theta}^T\dot{\Tilde{\theta}}.
\end{align}
On substituting (\ref{2ed2})  in (\ref{2vnd}), we have
\begin{align}\label{22vndd}
\dot{\mathcal{V}}_n={\mathcal{Q}}_n\left(\theta^{*T}\varphi+ u +\varepsilon _{n}-\frac{z_n}{\psi_n}\dot\psi_n\right)+{\tilde{\varepsilon}}_n\dot {\tilde{\varepsilon}}_n+\frac{1}{\lambda}\Tilde{\theta}^T\dot{\Tilde{\theta}}.
\end{align}

On substituting (\ref{2oaln}) in (\ref{22vndd}), we have
\begin{align}\label{2vndd}
\dot{\mathcal{V}}_n&={\mathcal{Q}}_n\theta^{*T}\varphi+{\mathcal{Q}}_nu+{\mathcal{Q}}_n\varepsilon _{n}-{\mathcal{Q}}_n\frac{z_i}{\psi_i}\dot\psi_n\\\nonumber&+{\tilde{\varepsilon}}_n\dot{\varepsilon} _n+\tilde\varepsilon _{n}k_{\varepsilon _n}\tilde\theta^{T}\varphi-k_{\varepsilon _n}\tilde\varepsilon _{n}^2+\frac{1}{\lambda}\Tilde{\theta}^T\dot{\Tilde{\theta}}.
\end{align}

Designing the control input and adaptive law  as
\begin{align}
u&=\mathcal{N}_n(\zeta_n)\alpha_n,~\text{where} \label{2vvn}\\
\dot \zeta_n&=\mathcal{Q}_n\alpha_n,~\text{and}\label{2gn}\\
\alpha_n&=k_nz_n+\hat\varepsilon_n+\frac{\mathcal{Q}_n}{2}-\frac{z_n}{\psi_n}{\dot \psi_n}+\hat{\theta}^T\varphi+\frac{\mathcal{Q}_n^{-1}k_{\varepsilon _n}^4}{8},\label{2an}\\
\dot{\hat \theta}&=\lambda\left(\mathcal{Q}_n\varphi- k_{\varepsilon_n}^2\hat \theta -\eta\hat \theta\right). \label{2upd}
\end{align}
where $\eta>0$ and $k_n>0$ are design parameters.\par
Using  (\ref{2vvn})-(\ref{2an}), (\ref{2vndd}) becomes
\begin{align}\label{2vndddd}
\dot {\mathcal{V}}_n&= -k_n\mathcal{Q}_nz_n+ \mathcal{N}_n(\zeta_n)\dot \zeta_n+\dot \zeta_n +\mathcal{Q}_n{\tilde{\varepsilon}}+{\tilde{\varepsilon}}_n\dot{\varepsilon} _n +\tilde\varepsilon _{n}k_{\varepsilon _n}\tilde\theta^{T}\varphi\notag\\
&\quad-\frac{\mathcal{Q}_n^2}{2}-k_{\varepsilon _n}\tilde\varepsilon _{n}^2 -\mathcal{Q}_n\tilde\theta^{T}\varphi-\frac{k_{\varepsilon _n}^4}{8}+\frac{1}{\lambda}\Tilde{\theta}^T\dot{\Tilde{\theta}}.
\end{align}

For further analysis, we need few inequality relations. They are as follows
\begin{enumerate}[i)]
	\item First term of (\ref{2vndddd}), i.e.,  $k_n\mathcal{Q}_nz_n$:\\
	Following (\ref{2Qi}) and using  Lemma 2,  we have
	\begin{align}
	- \frac{1}{2}\mathcal{Q}_nz_n=-\frac{1}{2}\frac{z_n^2}{\psi_n^2-z_n^2}\le- \frac{1}{2}\log \frac{\psi_n^2}{\psi_n^2-z_n^2} \label{2nnl}.
	\end{align}
	Multiplying  (\ref{2nnl}) on both sides by $2k_n$, we have  
	\begin{align}
	- k_n\mathcal{Q}_nz_n\le-2k_n \mathcal{L}_n \label{2Fn}.
	\end{align}
\end{enumerate}

\begin{enumerate}[i)]
	\setcounter{enumi}{1}
	\item Fourth, sixth, seventh and eighth term of (\ref{2vndddd}), i.e., $\mathcal{Q}_n{\tilde{\varepsilon}_n}+{\tilde{\varepsilon}}_n\dot{\varepsilon} _n+\tilde\varepsilon _{n}k_{\varepsilon _n}\tilde\theta^{T}\varphi-k_{\varepsilon _n}\tilde\varepsilon _{n}^2$.
\end{enumerate}

Following  Assumption 6 and (\ref{2bnd1}),  and  applying Young's inequality, we have
\begin{align}\label{2Tn}
\mathcal{Q}_n{\tilde{\varepsilon}_n}&+{\tilde{\varepsilon}}_n\dot{\varepsilon} _n-k_{\varepsilon _n}\tilde\varepsilon _{n}^2 +\tilde\varepsilon _{n}k_{\varepsilon _n}\tilde\theta^{T}\varphi\le\frac{\tilde \varepsilon_n^2}{2}+\frac{\mathcal{Q}_n^2}{2}+\frac{\tilde \varepsilon_n^2}{2}+\frac{{\bar \varepsilon}_n^2}{2}\\&-k_{\varepsilon_n}{\tilde{\varepsilon}}^2_n+\frac{1}{2}k_{\varepsilon _n}^2\norm{\tilde\theta}^2+\frac{1}{2}\tilde\varepsilon _{n}^2{\bar\varphi}^2,\notag\\
&=-\tilde \varepsilon_n^2\left(k_{\varepsilon_n}-1-\frac{{\bar\varphi}^2}{2}\right)+\frac{\mathcal{Q}_n^2}{2}+\frac{{\bar \varepsilon}_n^2}{2}+\frac{1}{2}k_{\varepsilon _n}^2\norm{\tilde\theta}^2.
\end{align}
\begin{enumerate}[i)]
		\setcounter{enumi}{2}
		\item For  the eleventh term of (\ref{2vndddd}), i.e. $\frac{1}{\lambda}{\Tilde{\theta}}^T\dot{\hat{\theta}}$.
	\end{enumerate}

Simplifying the expression $\frac{1}{\lambda}{\Tilde{\theta}}^T\dot{\hat{\theta}}$ using (\ref{2upd}), we have
	\begin{align}\label{2win}
	\frac{1}{\lambda}{\Tilde{\theta}}^T\dot{\hat{\theta}}&=\mathcal{Q}_n\Tilde{\theta}^T\varphi- k_{\varepsilon_n}^2\Tilde{\theta}^T\hat \theta-\eta\Tilde{\theta}^T\hat \theta
	\end{align}
	Using the inequality below
	\begin{align}
	-{\tilde \theta^T}{\hat{\theta}}\le\frac{1}{2}\left(\norm{\theta^*}^2-\norm{\tilde \theta}^2\right)
	\end{align}
	in (\ref{2win}), we have
	\begin{align}\label{2win1}
	\frac{1}{\lambda}{\Tilde{\theta}}^T\dot{\hat{\theta}}&\le \mathcal{Q}_n\Tilde{\theta}^T\varphi+\frac{1}{2}k_{\varepsilon_n}^2\norm{\theta^*}^2-\frac{1}{2}k_{\varepsilon_n}^2\norm{\tilde \theta}^2\notag\\
	&\quad+\frac{\eta}{2}\norm{\theta^*}^2-\frac{\eta}{2}\norm{\tilde \theta}^2.
	\end{align}
	Applying Young's inequality in the second term of  (\ref{2win1}), we have
	\begin{align}\label{2Fon}
	\frac{1}{\lambda}{\Tilde{\theta}}^T\dot{\hat{\theta}}&\le \mathcal{Q}_n\Tilde{\theta}^T\varphi+\frac{1}{8}k_{\varepsilon_n}^4+\frac{1}{2}\norm{\theta^*}^4\notag\\
	&\quad-\frac{1}{2}k_{\varepsilon_n}^2\norm{\tilde \theta}^2+\frac{\eta}{2}\norm{\theta^*}^2-\frac{\eta}{2}\norm{\tilde \theta}^2.
	\end{align}
	Using all the four inequalities (\ref{2Fn}),  (\ref{2Tn}), and (\ref{2Fon}) in  (\ref{2vndddd}), we have
\begin{align}\label{2222}
\dot {\mathcal{V}}_n&= -2k_n\mathcal{L}_n+ \mathcal{N}_n(\zeta_n)\dot \zeta_n+\dot \zeta_n -\tilde \varepsilon_n^2\left(k_{\varepsilon_n}-1-\frac{{\bar\varphi}^2}{2}\right)+\frac{{\bar \varepsilon}_n^2}{2}\notag\\
&\quad+\frac{1}{2}\norm{\theta^*}^4+\frac{\eta}{2}\norm{\theta^*}^2-\frac{\eta}{2}\norm{\tilde \theta}^2.
\end{align}
	The equation (\ref{2222}) can be further written as
	\begin{align}\label{2ldot1n}
	\dot {\mathcal{V}}_n&\le-\mu_n\mathcal{V}_n+  \mathcal{N}_n(\zeta_n)\dot \zeta_n+\dot \zeta_n +\varrho_n,
	\end{align}
	where  $\mu_n=\min\left(2k_n,2\left(k_{\varepsilon_n}-1-\frac{\bar{\varphi}^2}{2}\right),\lambda\eta\right) \text{and} ~ \varrho_n=\frac{{\bar \varepsilon}_n^2}{2}+\frac{1}{2}\norm{\theta^*}^4+\frac{\eta}{2}\norm{\theta^*}^2.$\par
 Following the same procedure as in the  step $1$, we can rewrite (\ref{2ldot1n}) as
\begin{align}\label{2vnd1}
	\mathcal{V}_n(t)&\le \mathcal{V}_n(0) + e^{-\mu_n t}\int_{0}^{t}{ \left({\mathcal{N}}_n(\zeta_n)+1\right) \dot\zeta_ne^{\mu_n\tau}d\tau}  +\frac{\varrho_n }{\mu_n} , 
	\end{align}
	In  (\ref{2vnd1}),   $\mathcal{V}_n(0)+{\varrho_n }/{\mu_n}$ is a constant. Let $c_n=\mathcal{V}_n(0)+{\varrho_n }/{\mu_n}$, then using Lemma 1 in (\ref{2vnd1}) we can say $\mathcal{V}_n(t),\zeta_n$ and $z_n,\hat \theta, \hat \varepsilon_n$ are  uniformly ultimately bounded.
	Due to the boundedness of   $z_n$, for $i=n-1$ in (\ref{2boundi}) we can say, the integral term $e^{-\mu_{n-1} t}\int_{0}^{t}{z_{n}^2e^{\mu_{n-1} \tau}d\tau}$ is bounded. Thus, based on  Lemma 1 and  (\ref{2ldot16}) for $i=n-1$ we can conclude that   $\mathcal{V}_{n-1}(t),\zeta_{n-1}$ and $z_{n-1}, \hat \varepsilon_{n-1}$ are  also uniformly ultimately bounded. Similarly, we can prove in that $\mathcal{V}_i(t),\zeta_i$ and $z_i, \hat \varepsilon_i$ are  uniformly ultimately bounded $\forall i \in \mathbb{N}_{n-2}$.
	
	\section{Boundedness and Convergence}\label{s2.5}
	\textbf{{Theorem 1:}} For a class of system (\ref{sys1}), under Assumptions 1-6
	and initial error condition $\abs{z_i(0)}<\abs{\psi(\bar x_i(0),0)}$, if the adaptive controller  is designed and  controller parameters are updated as given in (\ref{2vv1})-(\ref{2a1}), (\ref{2vvi})-(\ref{2ai}), (\ref{2vvn})-(\ref{2an})  and  (\ref{2upd}), respectively, then the closed-loop system  holds the listed properties:
	\begin{enumerate}[i)]
		\item All the signals are bounded.
		\item The system states will never contravene their respective constraints, i.e. $\abs{x_i}<\Psi_i(t)$.
		\item The closed-loop error signal $z_1$ will converge to a small neighbourhood of zero.
	\end{enumerate}
	\textbf{{Proof i).}} Following all the steps $1$ to $n$ of controller design and stability analysis, it is trivial to prove that all the signals in the closed-loop system are bounded.\par
	\textbf{{Proof ii).}} To prove this, we will use proof by contradiction. Let us assume that, for $i=1$ there exists some $t=\mathbb{T}$,  such that $\abs{z_1(\mathbb{T})}$ grows to $\psi(\mathbb{T})$. Then, substituting $\abs{z_{1}(\mathbb{T})}=\psi_1(\mathbb{T})$ in (\ref{2L}) makes  $\mathcal{L}_1=\frac{1}{2}\log\frac{\psi_1^2}{\psi_1^2-z_1^2}$ unbounded and based on  (\ref{2v1}),  $\mathcal{V}_1$ involve $\mathcal{L}_1$, i.e. $\mathcal{V}_1$ will becomes unbounded,  contradicting the previous proved results. Thus, for any $t$, $\abs{z_1(t)}<\psi_1(t)$. Similarly, we can prove this  $\forall i \in \{2, \hdots, n\}$. Hence, we have  
	\begin{align}
	\abs{z_i(t)}<\psi_i(t), ~~~\forall i\in  \mathbb{N}_n.\label{p1}    
	\end{align}
 As all the signals are bounded  $v_i \in L_\infty$, let $A_{i-1}=max\abs{v_{i-1}}$. From $x_i=z_i+v_{i-1}$ and $\abs{z_i}<\psi_i$, we have $\abs{x_i}<\abs{z_i}+\abs{v_{i-1}}<\psi_i+A_{i-1}$.  If $\psi_i=\Psi_i-A_{i-1}$ and design parameter are choosen to satisfy $-\Psi_i<A_{i-1}<\Psi_i$ then it is easy to know that      $\abs{x_i}<\Psi_i$. Then, the system state variables do not contravene their constraints.\par
  To make the controller design simple, we have not considered feasibility condition in controller design. In the next paper we will consider the feasibility condition.\par  
	\textbf{{Proof iii).}} Let $C_{\zeta_1}$ be the upper bound of integral term in (\ref{2ldot15}) 
	\begin{align}\label{2boundd}
	e^{-\mu_1 t}&\int_{0}^{t}{ \left({\mathcal{N}}_1(\zeta_1)+1\right) \dot\zeta_1e^{\mu_1\tau}d\tau}\notag\\
	&+e^{-\mu_1 t}\int_{0}^{t}{z_{2}^2e^{\mu_1 \tau}d\tau}\le C_{\zeta_1}.
	\end{align}
	Following (\ref{2v1}) and (\ref{2L}), and using (\ref{2boundd}),  we can write (\ref{2ldot15}) as
	\begin{align}\label{2inep}
	\frac{1}{2}\log\frac{\psi_1^2}{\psi_1^2-z_1^2}\le \mathcal{V}_1(t)&\le e^{-\mu_1 t}\left(\mathcal{V}_1(0)-\frac{\varrho_1}{\mu_1}\right)+\frac{\varrho_1 }{\mu_1}+C_{\zeta_1}.
	\end{align}
	On solving the above inequality, we have (\ref{2inep}) as
	\begin{align}\label{2inep1}
	\abs{z_1}\le\psi_1\sqrt{1-e^{-2\frac{\varrho_1 }{\mu_1}-2C_{\zeta_1}}e^{-2\left(\mathcal{V}_1(0)-\frac{\varrho_1}{\mu_1}\right)e^{-\mu_1 t}}}
	\end{align}
	For $t\rightarrow\infty$ in (\ref{2inep1}), we have 
	\begin{align}\label{2inep2}
	\abs{z_1}\le\psi_1\sqrt{1-e^{-2\frac{\varrho_1 }{\mu_1}-2C_{\zeta_1}}}.
	\end{align}
	In the above error bound of $z_1$, we can see that $z_1$ can be made arbitrarily small, by selecting the design parameters appropriately.\par
	
  \section{Simulation Results and Discussion}\label{s2.6}
  To show the effectiveness of proposed approach, it has been  applied to a nonlinear system  as given below. 
  \begin{equation}\label{seg2}
      \begin{split}
            \dot x_1&=x_2+d_1(t)\\
            \dot x_2&=-5x_1^3-2x_2+u+d_2(t)\\
            y&=x_1
        \end{split}
  \end{equation}
  	where $x_1$ and $x_2$ are the states, $u$ is  the control input, and $y=x_1$ is the output of the system. To verify the robustness of  proposed controller, disturbances $d_1=0.2\cos(\pi t)$ and $d_2=0.2\sin(\pi t)$ are considered in  the system.  Let, $y_d=\sin(t)$ be the desired output of system, and $\Psi_1=e^{-0.7t}+1.1$ and $\Psi_2=e^{-0.6t}+1.1$ be the constraints on system states $x_1$ and $x_2$, respectively. The goal is to design a control input $u$ such that the system output follows the desired trajectory $y_d$ and the system states do not contravene their respective constraints, i.e. $\abs{x_1}<\Psi_1$ and $\abs{x_2}<\Psi_2$.\par
The virtual controller and the sctual controller is designed as (\ref{2vv1})-(\ref{2a1}) and (\ref{2vvn})-(\ref{2an}), respectively. The weights of the RBF NN is updated using the (\ref{2upd}). The design parameter and initial values used in the simulation are:	$k_1=k_2=5$; $k_{\varepsilon_1}=k_{\varepsilon_2}=7$; $\eta=4$; $A_0=1$, $A_1=2$; $x_1(0)=0$, $x_2(0)=0$; $\lambda=14$, and $\zeta_1(0)=0$, $\zeta_2(0)=0$. The weights of the RBF NN are  chosen as a  $12 \times 1$ dimensional vector, where $12$ and $1$ represent the number of nodes in the hidden layer and the output of the NN, respectively.\par
\begin{figure}
     \centering
     \includegraphics[width=\linewidth, height=4cm]{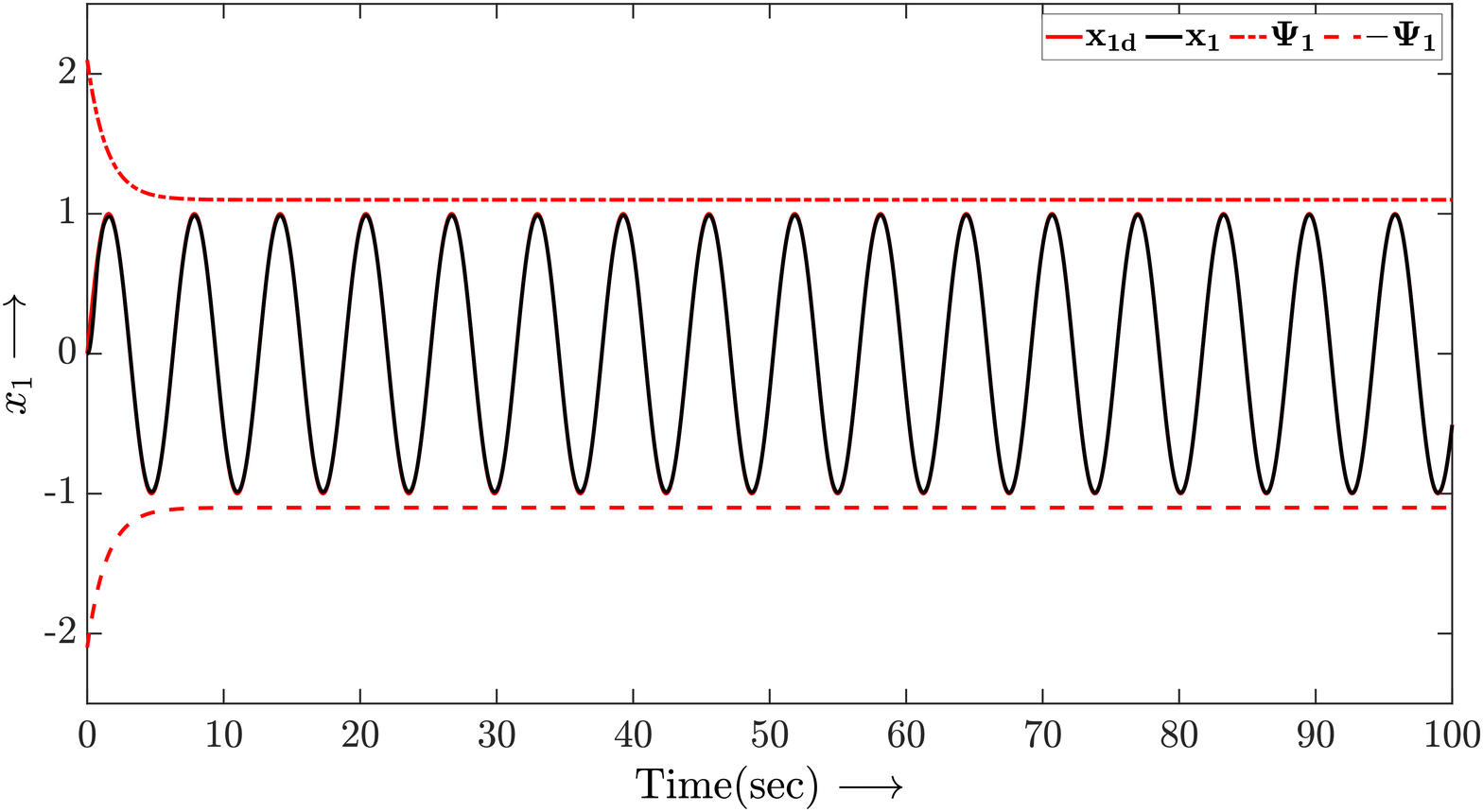}
     \caption{ Output tracking and boundedness performance of $x_1$.}
     \label{2x1}
     \centering
     \includegraphics[width=\linewidth, height=4cm]{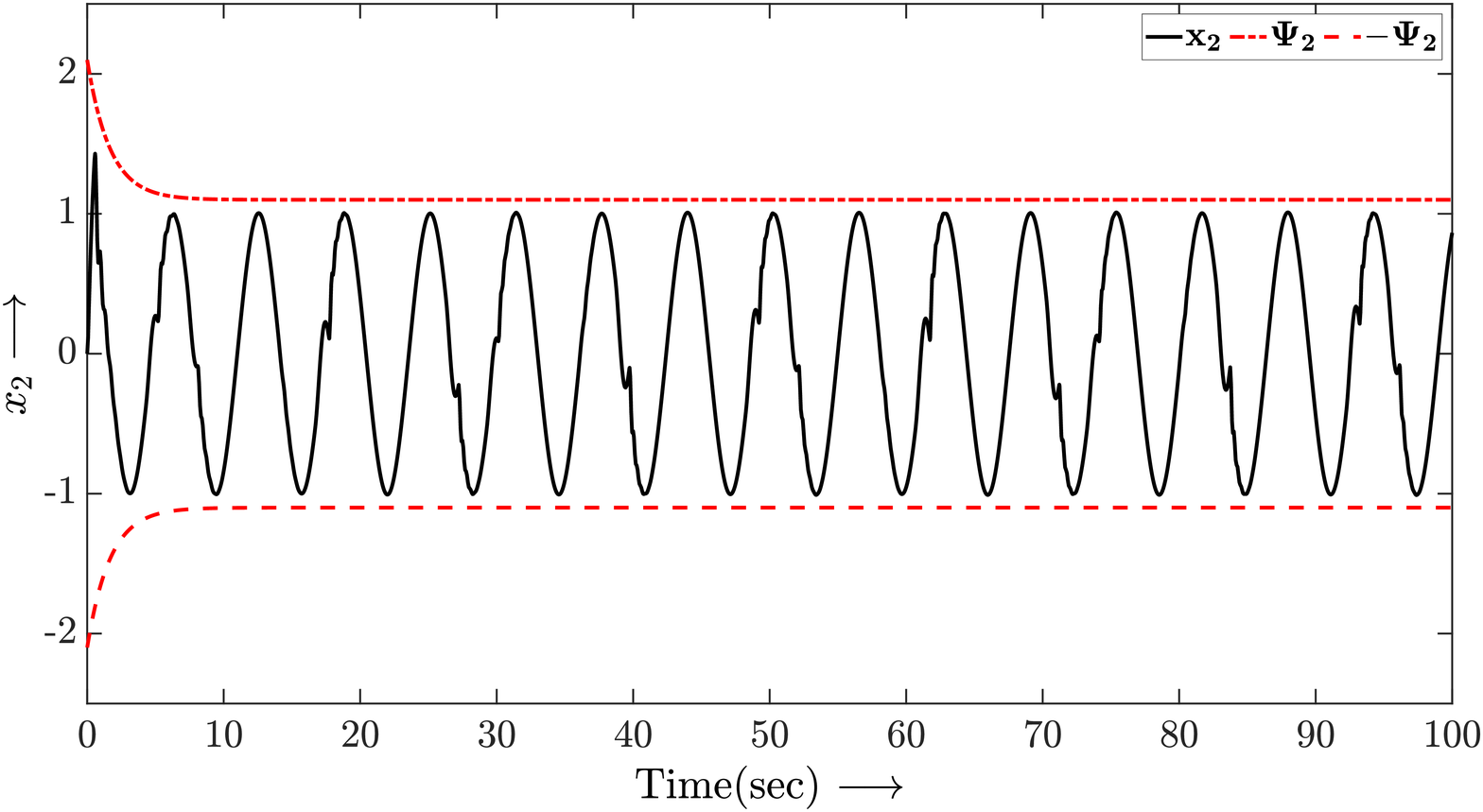}
     \caption{State trajectory and boundedness performance of $x_2$.}
     \label{2x2}
     \centering
     \includegraphics[width=\linewidth, height=4cm]{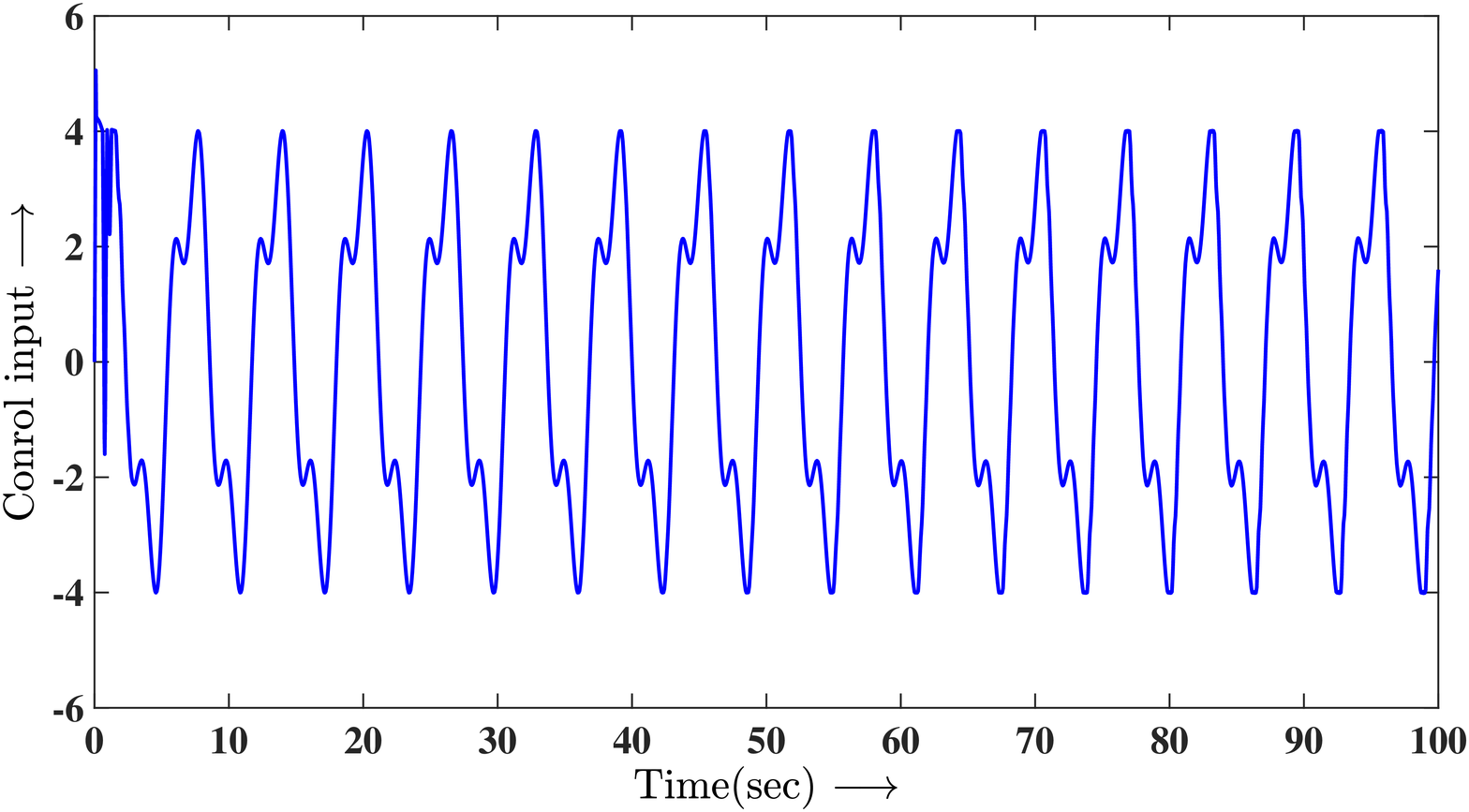}
     \caption{The control input signal.}
     \label{2u}
  \end{figure}

 Figs. \ref{2x1}-\ref{2x2} delineate the trajectories of the states and  its constraints. It can be seen that all the states are bounded in nature and do not contravene their respective constraints. Also, Fig. \ref{2x1} show that output tracks their reference effectively. Furthermore,  Figs. \ref{2x1}-\ref{2u} infer that all  signals in the closed-loop system are bounded in nature. 
  
  \section{Conclusion}\label{s2.7}
  A control strategy for a nonlinear system with symmetrical and time-varying state restrictions has been proposed. By implementing the proposed approach, no state violates its constraints, and the output follows the desired trajectory asymptotically. The simulation study validated the proposed control scheme's efficacy. 
  \ifCLASSOPTIONcaptionsoff
 \newpage
\fi
\bibliographystyle{ieeetr}
\bibliography{aa.bib}
\end{document}